\input harvmac

\def \om {\omega}

\def \k {\kappa} 
\def \F {{\cal F}}

\def \ha{{\textstyle{1\over 2}}}

\def \chi {\chi}

\def \l {\lambda}

\def \inv {^{-1}}
\def \ov {\over }

\def \lr { \lref}
\def\np {{  Nucl. Phys. }}
\def \pl {{  Phys. Lett. }}
\def \mpl {{ Mod. Phys. Lett. }}
\def \prl {{  Phys. Rev. Lett. }}
\def \pr  {{ Phys. Rev. }}

\def \cqg {{ Class. Quant. Grav. }}

\baselineskip8pt
\Title{
\vbox
{\baselineskip 6pt{\hbox{PUPT-1616}}{\hbox
{Imperial/TP/95-96/41}}{\hbox{hep-th/9604166}} {\hbox{
  }}} }
{\vbox{\centerline {Intersecting M-branes}
\vskip4pt
 \centerline { as Four-Dimensional Black Holes }
}}
\vskip -20 true pt
\medskip
\medskip
\centerline  { {  I.R. Klebanov\footnote {$^*$} {e-mail address:
 klebanov@puhep1.princeton.edu} }}

 \smallskip \smallskip

\centerline{\it Joseph Henry Laboratories }
\smallskip

\centerline{\it   Princeton University, Princeton, NJ 08544 }
\medskip
\centerline {and}

\medskip
\centerline{   A.A. Tseytlin\footnote{$^{\star}$}{\baselineskip8pt
e-mail address: tseytlin@ic.ac.uk}\footnote{$^{\dagger}$}{\baselineskip8pt
On leave  from Lebedev  Physics
Institute, Moscow.} }

\smallskip\smallskip
\centerline {\it  Theoretical Physics Group, Blackett Laboratory}
\smallskip

\centerline {\it  Imperial College,  London SW7 2BZ, U.K. }

\bigskip
\centerline {\bf Abstract}
\medskip
\baselineskip10pt
\noindent
We present two 1/8 supersymmetric intersecting p-brane 
solutions of 11-dimensional supergravity which upon
compactification to four dimensions reduce to extremal dyonic 
black holes with finite area of horizon. 
The first solution is a configuration  
of three intersecting 5-branes with an extra momentum flow 
along the common string. The second  describes
a system of two 2-branes and two 5-branes. Related 
(by compactification and T-duality) solution  of type IIB 
theory corresponds to a completely symmetric configuration of 
four intersecting 3-branes.
We suggest methods for counting the BPS degeneracy of
three intersecting 5-branes which, in the macroscopic limit,
reproduce the Bekenstein-Hawking entropy.

\medskip
%%%%%%%%%%%%%%%%%%%%%%%%%%%%%%%%%%%%%%%%%%%%%%%%%%%%%%%%%
\Date {April 1996}
%%%%%%%%%%%%%%%%%%%%%%%%%%%%%%%%%%%%%%%%%%%%%%%%%%%%%%%%%%%%%%%%%%%
\noblackbox
\baselineskip 14pt plus 2pt minus 2pt
%\baselineskip 20pt plus 2pt minus 2pt
%%%%%%%%%%%%%%%%%%%%%%%%%%%%%%%%%%%%%%%%%%%
\lr\MalLen{J. Maldacena and L. Susskind, hep-th/9604042.}
\lr\DasMat{S. Das and S. Mathur, hep-th/9601152.}
\lr \dgh {A. Dabholkar, G.W. Gibbons, J. Harvey and F. Ruiz Ruiz,  \np
B340 (1990) 33;
A. Dabholkar and  J. Harvey,  \prl
63 (1989) 478.
}
\lr\mon{J.P. Gauntlett, J.A. Harvey and J.T. Liu, \np B409 (1993) 363.}
\lr\chs{C.G. Callan, J.A. Harvey and A. Strominger, 
\np { B359 } (1991)  611.}
% in {\it 
%Proceedings of the 1991 Trieste Spring School on String Theory and
%Quantum Gravity}, J.A. Harvey {\it et al.,}  eds. (World Scientific, 
%Singapore 1992).}

\lr \CM{ C.G. Callan and  J.M.  Maldacena, 
  hep-th/9602043.} 
\lr\SV {A. Strominger and C. Vafa,   hep-th/9601029.}

\lr\MV {J.C. Breckenridge, R.C. Myers, A.W. Peet  and C. Vafa, HUTP-96-A005,  hep-th/9602065.}
\lr\vijay{V. Balasubramanian and F. Larsen, hep-th/9604189.}
\lr \CT{M. Cveti\v c and  A.A.  Tseytlin, 
\pl { B366} (1996) 95, hep-th/9510097. 
}
\lr \CTT{M. Cveti\v c and  A.A.  Tseytlin, \pr D53 (1996) 5619, 
 hep-th/9512031. 
}
\lr\LW{ F. Larsen  and F. Wilczek, 
  hep-th/9511064.    }
\lr\TT{A.A. Tseytlin, \mpl A11 (1996) 689,  hep-th/9601177.}
\lr \HT{ G.T. Horowitz and A.A. Tseytlin,  \pr { D51} (1995) 
2896, hep-th/9409021.}
\lr\khu{R. Khuri, \np B387 (1992) 315; \pl B294 (1992) 325.}
\lr\CY{M. Cveti\v c and D. Youm,
 UPR-0672-T, hep-th/9507090; UPR-0675-T, hep-th/9508058; 
  \pl { B359} (1995) 87, 
hep-th/9507160.}

\lr \jons{J.H. Schwarz, hep-th/9510086. }
\lr\ght{G.W. Gibbons, G.T. Horowitz and P.K. Townsend, \cqg 12 (1995) 297,
hep-th/9410073.}
\lr\dul{M.J. Duff and J.X. Lu, \np B416 (1994) 301, hep-th/9306052. }
\lr\hst {G.T. Horowitz and A. Strominger, hep-th/9602051.}
\lr\dull{M.J. Duff and J.X. Lu, \pl B273 (1991) 409. }
\lr \guv{R. G\"uven, \pl B276 (1992) 49. }
\lr \gups {S.S. Gupser, I.R.   Klebanov  and A.W. Peet, 
hep-th/9602135.}
\lr \dus { M.J. Duff and  K.S. Stelle, \pl B253 (1991) 113.}

\lr\hos{G.T.~Horowitz and A.~Strominger, Nucl. Phys. { B360}
(1991) 197.}
\lr\teit{R. Nepomechi, \pr D31 (1985) 1921; C. Teitelboim, \pl B167 (1986) 69.}
\lr \duf { M.J. Duff, P.S. Howe, T. Inami and K.S. Stelle, 
\pl B191 (1987) 70. }
\lr\duh {A. Dabholkar and J.A. Harvey, \prl { 63} (1989) 478;
 A. Dabholkar, G.W.   Gibbons, J.A.   Harvey  and F. Ruiz-Ruiz,
\np { B340} (1990) 33. }
\lr\mina{M.J. Duff, J.T. Liu and R. Minasian, 
\np B452 (1995) 261, hep-th/9506126.}
\lr\dvv{R. Dijkgraaf, E. Verlinde and H. Verlinde, hep-th/9603126;
hep-th/9604055.}
\lr\gibb{G.W. Gibbons and P.K. Townsend, \prl  71
(1993) 3754, hep-th/9307049.}
\lr\town{P.K. Townsend, hep-th/9512062.}
\lr\kap{D. Kaplan and J. Michelson, hep-th/9510053.}
\lr\hult{
C.M. Hull and P.K. Townsend, Nucl. Phys. { B438} (1995) 109;
P.K. Townsend, Phys. Lett. {B350} (1995) 184;
E. Witten, \np B443 (1995) 85; 
J.H. Schwarz,  \pl B367 (1996) 97, hep-th/9510086, hep-th/9601077;
P.K. Townsend, hep-th/9507048;
M.J. Duff, J.T. Liu and R. Minasian, 
\np B452 (1995) 261, hep-th/9506126; 
K. Becker, M. Becker and A. Strominger, Nucl. Phys. { B456} (1995) 130;
I. Bars and S. Yankielowicz, hep-th/9511098;
P. Ho{\v r}ava and E. Witten, Nucl. Phys. { B460} (1996) 506;
E. Witten, hep-th/9512219.}
\lr\beck{
K. Becker and  M. Becker, hep-th/9602071.}
\lr\aar{
O. Aharony, J. Sonnenschein and S. Yankielowicz, hep-th/9603009.}
\lr\ald{F. Aldabe, hep-th/9603183.}
\lr\ast{A. Strominger, hep-th/9512059.}
\lr \ttt{P.K. Townsend, hep-th/9512062.}
\lr \papd{G. Papadopoulos and P.K. Townsend, hep-th/9603087.}
\lr\jch {J. Polchinski, S. Chaudhuri and C.V. Johnson, 
hep-th/9602052.}
\lr \ddd{E. Witten, hep-th/9510135;
M. Bershadsky, C. Vafa and V. Sadov, hep-th/9510225;
A. Sen, hep-th/9510229, hep-th/9511026;
C. Vafa, hep-th/9511088;
M. Douglas, hep-th/9512077. }

\lr \gig{G.W. Gibbons, M.J. Green and M.J. Perry, 
hep-th/9511080.}

\lr \dufe{M.J. Duff, S.  Ferrara, R.R. Khuri and 
J. Rahmfeld, \pl B356 (1995) 479,  hep-th/9506057.}

\lr\stp{H. L\" u, C.N. Pope, E. Sezgin and K.S. Stelle, \np B276 (1995)  669, hep-th/9508042.}
\lr \duff { M.J. Duff and J.X. Lu, \np B354 (1991) 141. } 
\lr \pol { J. Polchinski, \prl 75 (1995) 4724,  hep-th/9510017.} 
\lr \iz { J.M. Izquierdo, N.D. Lambert, G. Papadopoulos and 
P.K. Townsend,  \np B460 (1996) 560, hep-th/9508177. }

\lr \US{M. Cveti\v c and  A.A.  Tseytlin, 
\pl {B366} (1996) 95, hep-th/9510097.  
}
\lr\mast{J.M. Maldacena and A. Strominger, hep-th/9603060.}
\lr \CY{M. Cveti\v c and D. Youm,
 \pr D53 (1996) 584, hep-th/9507090.  }
 \lr\kall{R. Kallosh, A. Linde, T. Ort\' in, A. Peet and A. van Proeyen, \pr { D}46 (1992) 5278.} 
\lr \grop{R. Sorkin, Phys. Rev. Lett. { 51 } (1983) 87;
D. Gross and M. Perry, Nucl. Phys. { B226} (1983) 29. 
}
\lr \myers{C. Johnson, R. Khuri and R. Myers, hep-th/9603061.} 

\lr \myk{R.R. Khuri and R.C. Myers, hep-th/9512061.}

\lr\lup {H. L\" u and C.N. Pope, hep-th/9512012; hep-th/9512153.}

\lr \KT{I.R. Klebanov and A.A. Tseytlin, hep-th/9604089 (revised).}

\lr \CYY{M. Cveti\v c and D. Youm, unpublished.}
\lr \CS{M. Cveti\v c and  A. Sen, unpublished.}
\lr \AT{ A.A. Tseytlin, hep-th/9604035.}
\lr \pap{G. Papadopoulos, hep-th/9604068.}

\lr \green{M.B. Green and M. Gutperle, hep-th/9604091.}
\lr \sen{A. Sen, \mpl  { A10} (1995) 2081, 
 hep-th/9504147. }
\lr \gib{G.W.  Gibbons and K. Maeda, \np {B298} (1988) 741.} 
\lr \dabb{  A. Dabholkar, J.P. Gauntlett, J.A. Harvey and D. Waldram, 
%``Strings as solitons and black holes as strings",
  hep-th/9511053.   }

\lr\mina{M.J. Duff, J.T. Liu and R. Minasian, 
\np B452 (1995) 261, hep-th/9506126.}

\lr\polch{
J.~Polchinski, Phys. Rev. Lett. 75 (1995)
4724,  hepth/9510017.}

%%%%%%%%%%%%%%%%%%%%%%%%%%%%%%%%%%%%%%%%%%%%%%%%%%%%%%%%%%%%%%%%%%%%%%%%%%
%%%%%%%%%%%%%%%%%%%%%%%%%%%%%%%%%%%%%%%%%%%%%%%%%%%%%%%%%%%%%%%%%%%%%%%%%%%%
\newsec{Introduction}
%%%%%%%%%%%%%%%%%%%%%%%%%%%%%%%%
The existence of 
supersymmetric extremal dyonic black holes with finite area of the  horizon 
provides a possibility   of a statistical  
understanding \sen\ of the Bekenstein-Hawking entropy from the point of view of string theory \refs{\LW,\CTT,\SV}.
Such  black hole solutions are found  
in four  \refs{\kall,\CY,\US}  and  five  \refs{\SV,\TT} 
dimensions but not in $D >5$ \refs{\KT,\CYY}. 
While the D-brane BPS state counting derivation of the entropy is relatively 
straightforward for the $D=5$  black holes \refs{\SV,\CM}, 
it is  less transparent  in the $D=4$ case,
a complication being the presence of a 
solitonic 5-brane  or Kaluza-Klein monopole 
in addition to a  D-brane configuration in the descriptions 
used in \refs{\mast,\myers}.

One may hope to find a different lifting of the dyonic $D=4$
black hole to $D=10$ string theory that may  correspond to 
a purely D-brane configuration.
 A related question is about the embedding of the $D=4$ dyonic black holes 
into $D=11$ supergravity (M-theory)  which would allow to reproduce 
their  entropy by  counting  the  corresponding BPS states
using the M-brane approach similar to the one 
applied in the  $D=5$ black hole case  
in \dvv.

 As was found in \AT,  the (three-charge, finite area) 
$D=5$ extremal black hole
can be  represented in M-theory by a configuration of orthogonally intersecting 
2-brane and 5-brane (i.e. $2\bot 5$)
with a momentum 
flow along the common string, or
by configuration of three 2-branes intersecting  over a point
($2\bot2\bot2$).
A particular   embedding of (four-charge, finite area) $D=4$ black hole 
into $D=11$ theory given in \AT\ 
can be  interpreted as a similar $2\bot 5$  configuration 
 `superposed' with a Kaluza-Klein monopole.

Below we shall demonstrate that it is possible to  get rid of the complication 
associated with  having the Kaluza-Klein monopole.   
There exists  a  simple 1/8 supersymmetric 
configuration of {\it four}
intersecting M-branes ($2\bot 2\bot 5\bot 5$)
with  diagonal $D=11$ metric.  
Upon compactification along six  isometric directions
it reduces to the dyonic $D=4$ black hole  with
finite area and all scalars being regular at the horizon.

The corresponding $2\bot 2\bot 4\bot 4$
 solution of type IIA  $D=10$ superstring theory (obtained by 
dimensional reduction along a direction common to the two 5-branes) 
is $T$-dual to a $D=10$ solution of type IIB    theory which 
describes
a remarkably symmetric  configuration of {\it four}   intersecting 
3-branes.\foot{Similar D-brane configuration was 
discussed in \refs{\vijay,\green}.
Note that it is a combination  of
 {\it four} and not  {\it three} intersecting 
3-branes  that is related (for the special
choice of equal charges)  
to the non-dilatonic ($a=0$) RN $D=4$ black hole. 
$T$-dual configuration of one 0-brane and  three intersecting 
4-branes of type IIA 
theory was considered  in \CS.}

Our discussion  will follow closely that of \AT\ 
where an  approach  to constructing intersecting  supersymmetric
p-brane solutions (generalising that of \papd)
  was presented.\foot{ Intersecting p-brane solutions in \refs{\papd,\AT}
and below are isometric in all directions internal to all constituent p-branes
(the background fields depend only on the remaining common
transverse directions). They are different from possible
 virtual configurations
where, e.g., a (p-2)-brane ends  (in transverse  radial direction)
on a p-brane \ast. A configuration of
   p-brane  and  p$'$-brane  intersecting in  p+p$'$-space
 may be also considered as a  special anisotropic   p+p$'$-brane.
 There  may  exist  
more general solutions (with constituent p-branes effectively
having different transverse
spaces \refs{\papd,\pap})  which  may `interpolate'
between intersecting p-brane solutions
and solutions with one p-brane  ending on another  in
the transverse direction of the latter.}
The 
supersymmetric configurations of two  or three   intersecting 2- and 5-branes  of $D=11$ supergravity
which preserve  1/4 or 1/8  of maximal supersymmetry are  
 $2\bot 2,\ 5\bot 5,\ 2\bot 5,$ $\ 2\bot 2 \bot 2,\ 5 \bot 5 \bot 5,\  2 \bot 2 \bot 5$ and  $ \ 2  \bot 5 \bot 5$. 
Two 2-branes can intersect over a point, two 5-branes -- over a 3-brane
(which in turn can intersect over a string),  2-brane and  5-brane
can intersect over a string \papd.
There exists  a simple `harmonic function' 
rule  which governs the construction of 
composite supersymmetric p-brane solutions in both $D=10$ and $D=11$: 
 a separate  harmonic function is assigned to each constituent  
1/2 supersymmetric p-brane.

Most of  the  configurations with 
{\it four} intersecting M-branes, namely, 
  $2\bot 2\bot 2\bot 2 $, \  $ 2\bot 2\bot 2\bot 5$ and  
$5\bot 5\bot 5\bot 2 $  are 1/16 supersymmetric and 
 have   transverse $x$-space dimension equal to two
($5\bot 5\bot 5\bot 5$ configuration with 5-branes intersecting 
over 3-branes to preserve supersymmetry does not fit into 11-dimensional 
space-time). 
 Being described in terms 
of  harmonic functions  of $x$ they are thus 
  not asymptotically flat in transverse directions.
There exists, however, a remarkable  exception  -- the 
configuration   $2\bot 2 \bot 5\bot 5$ 
which  (like   $5\bot 2\bot 2$,\ $5\bot 5\bot 2$ and 
$5\bot 5\bot 5$) has 
 transverse dimension equal to three 
and the fraction of unbroken supersymmetry equal to  1/8 (Section 3).
Upon compactification to $D=4$  it reduces 
to the extremal dyonic  black hole  with {\it four}  different charges  
and finite area of the horizon. 

Similar $D=4$ black hole background can be obtained also from  
 the  the `boosted' version of  the  $D=11$ 
$5\bot 5\bot 5$   solution \AT\ (Section 2)\foot{The `boost' along the common string corresponds to a Kaluza-Klein electric charge part in the $D=11$ metric
which is `dual' to a Kaluza-Klein monopole part present in the $D=11$ embedding 
of dyonic black hole in \AT.}
as well from the $3\bot 3\bot 3\bot 3$ solution of $D=10$   type IIB theory (Section 4). 
The two $D=11$ configurations   
$5\bot 5\bot 5+$`boost' and $2\bot 2 \bot 5\bot 5$  
reduce in $D=10$  to  $0\bot 4\bot 4\bot 4$  and $2\bot 2 \bot 4\bot 4$
solutions of $D=10$ type IIA theory which are related by $T$-duality.

In Section 5 we shall suggest methods for counting the BPS entropy 
of three intersecting 5-branes which reproduce the Bekenstein-Hawking
entropy of the $D=4$ black hole. This seems to explain the microscopic
origin of the entropy directly in 11-dimensional terms.

\def \F {{\cal F}}

%%%%%%%%%%%%%%%%%%%%%%%%%%%%%%%%%%%%%%%%%%%%%%%%%%%%%%%%%%%%%%%
\newsec{`Boosted'  $5\bot 5\bot 5$  solution of $D=11$ theory}
%%%%%%%%%%%%%%%%%%%%%%%%%%%%%%%%%%%%%%%%%%%%%%%%%%%%%%%%%%%
The $D=11$ background  corresponding to   $5\bot 5\bot 5$ configuration \papd\  is \AT\ 
 \eqn\fivv{
d s^2_{11} =  (F_1 F_2 F_3)^{-2/3}
 \big[ F_1 F_2 F_3 (-  dt^2  + dy_1^2)  } $$
+ \  F_2 F_3 (dy^2_2 + dy_3^2)  + 
  F_1 F_3 (dy^2_4 + dy_5^2)  +  F_1 F_2 (dy^2_6 + dy^2_7)  
+  dx_s dx_{s}\big] \ ,     $$
\eqn\ffrof{ {\F_4}
 =3( *dF\inv_1 \wedge dy_2\wedge dy_3  +  *dF\inv_2 \wedge dy_4\wedge dy_5 +  *dF\inv_3 \wedge dy_6\wedge dy_7 ) \ .  }
Here $\F_4$ is the 4-form field strength and $F_i$ are the inverse powers of harmonic functions of $x_s \ (s=1,2,3)$. 
 In the simplest 1-center case discussed below 
 $F_i^{-1} = 1 + P_i/r$ \ $(r^2=x_sx_s$). 
The $*$-duality  is defined with respect to the transverse  3-space. 
 The coordinates $y_n$ internal to the three 5-branes can be  identified 
according to  the $F_i$ factors inside the square brackets in the metric:
$(y_1,y_4,y_5,y_6,y_7)$ belong to the first 5-brane, 
$(y_1,y_2,y_3,y_6,y_7)$  to the second and 
$(y_1,y_2,y_3,y_4,y_5)$  to the third. 
5-branes  intersect over three 3-branes which in turn intersect over a common string along $y_1$. 
If $F_2=F_3=1$ the above  background reduces to the  
single 5-brane solution \guv\ 
with the harmonic function $H=F_1\inv$ independent of the two of transverse 
coordinates (here $y_{2}, y_{3}$).
The case of $F_3=1$ describes  two 
5-branes   intersecting  over a 3-brane.\foot{The corresponding 
1/4 supersymmetric 
background also has 3-dimensional transverse space and reduces 
to a $D=4$ black hole with two charges (it has $a=1$ black hole
metric when  two charges are equal). The $5\bot 5$ 
configuration compactified to $D=10$ gives $4\bot 4$
solution of type IIA theory which is $T$-dual to 
$3\bot 3$ solution of type IIB theory.}
The special case of $F_1=F_2=F_3$  is   the solution 
found in \papd.

Compactifying $y_1,..,y_7$ on circles we learn that the effective `radii' (scalar moduli fields in $D=4$)
behave regularly both at $r=\infty$  and at $r=0$ with the exception
of 
 the `radius' of $y_1$. It is possible to  stabilize  the corresponding scalar 
 by  adding a `boost' along 
 the common string. The metric of the resulting more general solution  \AT\ is 
(the expression for $\F_4$ remains the same)  
 \eqn\fivve{
d s^2_{11} =  (F_1 F_2 F_3)^{-2/3}
 \big[ F_1 F_2 F_3 ( dudv  +  K du^2)  } $$
+ \  F_2 F_3 (dy^2_2 + dy_3^2)  + 
  F_1 F_3 (dy^2_4 + dy_5^2)  +  F_1 F_2 (dy^2_6 + dy^2_7) 
+  dx_s dx_{s}\big] \ .     $$ 
Here $u=y_1-t, \ v=2t$ and $K$  is a 
 harmonic function of the three coordinates $x_s$. 
A non-trivial $K =1+  Q/r$ 
describes a momentum flow along the string  direction.\foot{The metric \fivve\
with $F_i=1$ (i.e. $ds^2= -K\inv dt^2 + K [dy_1  + (K\inv -1)dt]^2 + dy_ndy_n + dx_sdx_s$) reduces  upon  compactification  along $y_1$ direction to 
 the $D=10$ type IIA  R-R 0-brane background \hos\ with
$Q$ playing the role of the KK electric charge.}

$Q$ also has an interpretation of  
The  $D=11$ metric \fivve\  is regular at the $r=0$ horizon 
and has a {\it non-zero}  9-area of the horizon 
 (we assume that all $y_n$ have period $L$)
\eqn\are{ A_9 = 4\pi L^7 [r^2  {K^{1/2}(F_1F_2F_3)^{-1/2}}]_{r\to 0}=
4\pi L^7  \sqrt {Q P_1P_2P_3} \ . } 
Compactification along $y_2,...y_7$ leads to a solitonic 
$D=5$ string. Remarkably, the corresponding 6-volume is constant
so that one gets directly the Einstein-frame metric
\eqn\soo{ ds^2_5 = H\inv (dudv + Kdu^2) + H^{2} dx_sdx_s \ , \ \ \ \ 
H\equiv (F_1 F_2 F_3)^{-1/3} \ . } 
Further compactification along $y_1$ or $u$ gives the  
 $D=4$ (Einstein-frame) 
metric which is  
isomorphic to the one of the dyonic black hole  \CY\ 
\eqn\ddd{ ds^2_4 =- \l (r)  dt^2 + \l\inv (r)  (dr^2 + r^2
d\Omega^2_2) \ ,  }
\eqn\luo{
 \l  (r) = \sqrt{K\inv  F_1F_2F_3  } = 
 {r^2\ov \sqrt
 {(r + {Q }) (r + {P_1})(r + {P_2})(r + {P_3})} }\ . 
}
Note, however, that  in 
 contrast to the  dyonic black hole background of \refs{\CY} which has 
two electric and  two magnetic charges
here there is  one electric (Kaluza-Klein) and  3 magnetic 
 charges. From the $D=4$ point of view the two backgrounds  
are related by $U$-duality. 
 The corresponding 2-area of the $r=0$ horizon is of course  $A_9/L^7$.

In the special case when all 4 harmonic functions are equal 
($K=F_i=H\inv$) 
the metric \fivve\ becomes
\eqn\vve{
d s^2_{11} =  H\inv dudv  +  du^2  
+ dy^2_2 + ... + dy_6^2  
+ H^2 dx_s dx_{s} \  }
$$ = \ - H^{-2} dt^2 +  H^2 dx_s dx_{s} + [dy_1 + (H\inv-1) dt]^2 +  dy^2_2 + ... + dy_6^2 \ , $$
and corresponds  to a charged solitonic string in $D=5$ or 
 the Reissner-Nordstr\"om  ($a=0$) black hole in $D=4$
(`unboosted'  $5 \bot 5\bot 5$  solution with $K=1$ and equal $F_i$
reduces   to $a=1/\sqrt 3$ dilatonic $D=4$ black hole \papd).

A compactification of this $5\bot 5\bot 5+$`boost' configuration to $D=10$
along $y_1$ gives a type IIA solution corresponding to three 
4-branes intersecting over 2-branes plus additional  Kaluza-Klein 
(Ramond-Ramond vector)
electric charge  background, or, equivalently, 
to the  $0\bot 4 \bot 4 \bot 4$  configuration of three 4-branes 
intersecting over 2-branes which in turn intersect over  a  $0$-brane. 
 If instead we compactify along a direction 
common only to two of the three 5-branes we get $4\bot 4\bot 5+$`boost'
type IIA solution.\foot{This may be compared to  another  type IIA configuration (consisting of solitonic 5-brane
lying  within a  
R-R 6-brane, both being   intersected 
over a `boosted' string 
by  a R-R 2-brane) which also reduces  \refs{\mast,\AT} 
 to the dyonic $D=4$ black hole.}
 Other related solutions of type IIA and IIB 
theories can be obtained by applying $T$-duality  and $SL(2,Z)$ duality.
 
%%%%%%%%%%%%%%%%%%%%%%%%%%%%%%%%%%%%%%%%%%%%%%%%%%%%%%
\newsec{ $2\bot 2 \bot 5\bot 5$ \  solution of $D=11$ theory}
%%%%%%%%%%%%%%%%%%%%%%%%%%%%%%%%%%%%%%%%%%%%%%%%%%%%%%%%%%%

Solutions with four intersecting M-branes are constructed 
according to the  rules  discussed   in \AT.
The $ 2\bot 2 \bot 5\bot 5$ configuration is described by the 
following background
\eqn\fiv{
d s^2_{11} =  (T_1T_2)^{-1/3} (F_1 F_2)^{-2/3}
 \big[ - T_1 T_2 F_1 F_2  \ dt^2  } $$
+ \ T_1 F_1  dy_1^2 +  T_1 F_2  dy_2^2    
 + T_2 F_1  dy_3^2 +  T_2  F_2  dy_4^2
+  F_1F_2 ( dy_5^2 + dy_6^2 + dy_7^2) 
+  dx_s dx_{s}\big] \ ,     $$
\eqn\ffd{ {\F_4}
 = -3dt\wedge (dT_1\wedge  dy_1\wedge d y_2 +
  dT_2\wedge  dy_3\wedge d y_4     )}
$$ +\ 
3( *dF\inv_1 \wedge dy_2 \wedge dy_4 +  *dF\inv_2 
\wedge dy_1 \wedge dy_3 ) \  . $$
Here $T_i\inv $  are harmonic functions corresponding 
to the 2-branes and $F_i\inv $ are harmonic functions corresponding 
to the 5-branes, i.e.
\eqn\func{ T_i\inv = 1 + {Q_i\ov r} \ , 
\ \ \ \ \  \ F_i\inv = 1 + {P_i \ov  r} \ . } 
$(y_1,y_2)$  belong to  the first and $(y_3,y_4)$ to the  second 2-brane. 
$(y_1,y_3,y_5,y_6,y_7)$  and 
$(y_2,y_4,y_5,y_6,y_7)$ are the coordinates 
of the two 5-branes. Each 2-brane  intersects each  5-brane over a string.
2-branes intersect over a 0-brane ($x=0$) 
and 5-branes intersect over a 3-brane.

Various special cases include, in particular,  the  2-brane solution \dus\ 
($T_2=F_1=F_2=1$), as well as $5\bot 5$ ($T_1=T_2=1$) \papd\ and  
$2\bot 5$  ($T_1=F_2=1$), $2\bot 2\bot 5$ ($F_2=1$), 
$2\bot 5\bot 5$ ($T_2=1$) \AT\ configurations   
(more precisely, their 
limits when the harmonic functions do not depend on a number 
of transverse coordinates). 

As in the case of the $5\bot 5\bot 5+$`boost' solution \fivve,\ffrof,  
the metric \fiv\ is regular at the  $r= 0$ horizon
(in particular, all internal $y_n$-components smoothly interpolate 
between  finite values at $r\to \infty$ and $r\to 0$)
with the 9-area of the horizon being (cf.\are)
\eqn\area{ A_9 = 4\pi L^7 [r^2 {(T_1T_2 F_1F_2 )^{-1/2} }]_{r\to 0}=
4\pi L^7  \sqrt {Q_1Q_2 P_1P_2} \ . } 
The compactification of $y_n$ on 7-torus    
leads to a  $D=4$  background with the 
metric  which is again   the dyonic black hole one \ddd, now with  
\eqn\lel{  \l  (r) = \sqrt{ T_1 T_2 F_1F_2} = 
 {r^2\ov \sqrt
 {(r + {Q_1})(r + {Q_2})(r + {P_1})(r + {P_2 })} }\ . 
}
In addition, there are 
 two electric and two magnetic vector fields   (as in \CY) 
and  also 7 scalar  fields. The two electric and two magnetic 
charges are {\it directly} 
 related to the 2-brane and 5-brane charges (cf. \ffd).

When all 4 harmonic functions are equal 
($T_i\inv=F_i\inv=H$) 
the metric \fiv\ becomes (cf. \vve)
\eqn\vte{
d s^2_{11} =  - H^{-2} dt^2   
+ H^2 dx_s dx_{s}+ dy^2_1 + ... + dy_7^2   \ , }
i.e. describes  a direct product of a   $D=4$ 
 Reissner-Nordstr\" om  black hole  and a 7-torus.

Thus there  exists an 
 embedding of the dyonic $D=4$ black holes
into $D=11$ theory  which corresponds to a  remarkably symmetric 
combination of M-branes only. In contrast to the 
embeddings with a Kaluza-Klein monopole \AT\ or 
electric charge (`boost')  \fivve,\vve\ it has a 
 diagonal $D=11$ metric.

%%%%%%%%%%%%%%%%%%%%%%%%%%%%%%%%%%%%%%%%%%%%%%%%%%%%%%%%%%%%%%%%
\newsec{$3\bot 3\bot  3\bot 3$ solution of type IIB theory}
%%%%%%%%%%%%%%%%%%%%%%%%%%%%%%%%%%%%%%%%%%%%%%%%%%%%%%%%%%%%%%%% 
Dimensional reduction of the background \fiv,\ffd\ to $D=10$
along a direction common to the two  5-brane (e.g. $y_7$)
gives a type IIA theory solution representing the R-R p-brane 
configuration  $2\bot 2\bot 4\bot 4$.
This
configuration is  $T$-dual to
 $0\bot 4\bot 4\bot 4$  one which is the dimensional reduction 
of the  
$5\bot 5\bot 5+$`boost'  solution. This suggests also  a relation 
 between the two $D=11$ configurations discussed in Sections 2 and 3.

 $T$-duality along one of the two  directions common to 4-branes 
transforms $2\bot 2\bot 4\bot 4$ into
 the $3\bot 3\bot 3\bot 3$ solution of type IIB theory.
The explicit form of the latter can be found  also 
directly in $D=10$ type IIB theory (i.e. independently of the 
above $D=11$ construction)  using the method of  \AT,  
where 
the  1/4 supersymmetric solution corresponding to two  intersecting 
3-branes was given. 
One finds the following  $D=10$ metric and 
self-dual 5-form (other $D=10$ fields are trivial) 
 \eqn\fwww{
ds^2_{10}= (T_1T_2T_3T_4)^{-1/2}  
\big[  - T_1T_2T_3T_4\  dt^2   } $$
   + \  T_1T_2 dy_1^2  +   T_1T_3 dy_2^2 
+   T_1T_4 dy_3^2 +   T_2T_3 dy_4^2+   T_2T_4 dy_5^2+   T_3T_4 dy_6^2+  
 dx_s dx_{s}\big] \ ,     $$
\eqn\fqe{
\F_5= dt\wedge (dT_1 \wedge dy_1 \wedge dy_2 \wedge dy_3
+  dT_2 \wedge dy_1 \wedge dy_4 \wedge dy_5 } $$ 
+ \ dT_3 \wedge dy_2 \wedge dy_4 \wedge dy_6 
+ dT_4 \wedge dy_3 \wedge dy_5 \wedge dy_6)$$ 
$$ + \ *dT_1\inv \wedge dy_4 \wedge dy_5 \wedge dy_6
+ *dT_2\inv \wedge dy_2 \wedge dy_3 \wedge dy_6 $$ $$ 
+ \ *dT_3\inv \wedge dy_1 \wedge dy_3 \wedge dy_5
+ *dT_4\inv \wedge dy_1 \wedge dy_2 \wedge dy_4 \ . $$
The coordinates of the four 3-branes
are $(y_1,y_2,y_3)$, $(y_1,y_4,y_5)$, $(y_2,y_4,y_6)$ and 
$(y_3,y_5,y_6)$, i.e. each pair of 3-branes intersect over 
a string and all 6 strings intersect at one point.
$T_i$ are the inverse harmonic functions corresponding to each 
3-brane, $T_i\inv= 1 + Q_i/r$.
 Like the   $2\bot 2\bot 5\bot 5$  background 
of $D=11$ theory this $D=10$ solution is 1/8 supersymmetric, 
 has 3-dimensional transverse 
space   and diagonal $D=10$ metric. 

Its special cases include the single 3-brane \refs{\hos,\dull}
 with harmonic function independent of 3 of 6 transverse coordinates
($T_2=T_3=T_4=1$), $3\bot3$ solution found in \AT\ ($T_3=T_4=1$)
and also $3\bot3\bot3$ configuration ($T_4=1$). The  1/8 supersymmetric  
$3\bot3\bot3$ configuration also has 3-dimensional 
transverse space\foot{Similar configurations 
of three and  four intersecting 
3-branes, and, in particular,  their  invariance under 
the 1/8 fraction of maximal supersymmetry  were  discussed 
in $D$-brane representation in \refs{\green,\vijay}.} 
but the corresponding $D=10$ metric
 \eqn\fqw{
ds^2_{10}= (T_1T_2T_3)^{-1/2}  
\big[  - T_1T_2T_3 \  dt^2   } $$
   + \  T_1T_2 dy_1^2  +   T_1T_3 dy_2^2 
+   T_1 dy_3^2 +   T_2T_3 dy_4^2+   T_2 dy_5^2+   T_3 dy_6^2+  
 dx_s dx_{s}\big] \ ,     $$
 is singular at $r=0$  and has zero 
 area of the $r=0$ horizon.\foot{This is similar to what 
one finds for the `unboosted' $5\bot 5\bot 5$ configuration \fivv,\ffrof.
As is well-known from 4-dimensional 
point of view, one does need {\it four} charges to 
get a regular behaviour of scalars near the horizon and  finite area.}

As in the two $D=11$ cases discussed in the  previous sections,  
the metric of the  $3\bot3\bot3\bot 3$ solution  \fwww\
has $r=0$ as a regular horizon with  finite  8-area 
(cf.\are,\area)
\eqn\arr{ A_8 = 4\pi L^6 [r^2 {(T_1T_2 T_1T_2 )^{-1/2} }]_{r\to 0}=
4\pi L^6  \sqrt {Q_1Q_2 Q_3 Q_4} \ . } 
$A_8/L^6$ is  the area of the horizon of the corresponding dyonic 
$D=4$ black hole  with the metric 
\ddd\  and 
\eqn\lele{  \l  (r) = \sqrt{ T_1 T_2 T_3 T_4} = 
 {r^2\ov \sqrt
 {(r + {Q_1})(r + {Q_2})(r + {Q_3})(r + {Q_4})} }\ . 
}
The gauge field configuration here involves 4 pairs of equal electric and magnetic charges.  When all charges  are equal,  the $3\bot3\bot3\bot 3$ metric
\fwww\ compactified to $D=4$  reduces 
 to the $a=0$  black hole metric (while the 
$3\bot3\bot3$  metric \fqw\ reduces to the  $a=1/\sqrt 3 $ 
black hole metric \gib).

%%%%%%%%%%%%%%%%%%%%%%%%%%%%%
\newsec{Entropy of  $D=4$ Reissner-Nordstr\"om  black hole}
%%%%%%%%%%%%%%%%%%%%%%%%%%
Above we  have demonstrated the existence of 
supersymmetric extremal $D=11$ and $D=10$  configurations with finite entropy 
which are built solely out 
of the fundamental $p$-branes of the corresponding theories 
(the 2-branes and the 5-branes of the M-theory and the
3-branes of type IIB theory)
and reduce upon compactification to  $D=4$ dyonic black hole backgrounds 
with regular horizon.  

Namely, 
 there  exists an 
 emdedding of a  four dimensional  dyonic  black hole
(in particular, of the non-dilatonic Reissner-Nordstr\" om  black hole)
into $D=11$ theory  which corresponds to a  
combination of M-branes only.
This may allow an application of the   approach  similar to the one of  \dvv\
to the derivation of the entropy \area\ by counting the 
number of different BPS excitations of the  $2\bot 2\bot 5\bot 5$ 
 M-brane configuration.

The $3\bot 3\bot 3\bot 3$  configuration 
 represents an  
embedding 
of the  1/8 supersymmetric dyonic  
$D=4$ black  hole into type IIB superstring theory 
which is remarkable in that  all four charges
enter  symmetrically.
It is natural to expect that there should  exist
a microscopic counting of the BPS states 
which reproduces the Bekenstein-Hawking entropy  
in a ($U$-duality invariant) way that treats all four 
charges on an equal footing. 

Although we hope to eventually attain
a general understanding of this problem, 
in what follows we shall  discuss the counting
of BPS states for one specific example discussed above:
the M-theory configuration  \fivve,\ffrof\ 
of the three intersecting 5-branes
with a common line. 
Even though  the counting rules of M-theory
are not entirely clear, we see an advantage to doing this
from M-theory point of view as
compared to previous discussions in the context of 
string theory \refs{\mast,\myers}:   the
11-dimensional problem is more symmetric.
Furthermore, apart from
the entropy problem, we may learn something about the M-theory.

%%%%%%%%%%%%%%%%%%%%%%%%%%%%%%%%%%%%%%%%%%
\subsec{Charge quantization in M-theory and 
the Bekenstein-Hawking entropy}
%%%%%%%%%%%%%%%%%%%%%%%%%%%%%%%%%%%%%%%%%%%%%%%
Upon dimensional reduction to four dimensions, the boosted
$5\bot 5\bot 5$ solution \fivve, \ffrof,
reduces to the 4-dimensional
black hole with three  magnetic charges, $P_1$, $P_2$ and $P_3$,
and an electric charge $Q$. The electric charge
is proportional 
to the momentum along the intersection string   of 
length $L$,
$\ {\cal P} = 2\pi N/L$.
The general relation between the coefficient $Q$
in the harmonic function $K$ appearing in \fivve\ and the momentum 
along the 
$D=5$ string (cf.\soo)  wound 
around a compact dimension of length $L$ is (see e.g. \dabb) 
\eqn\qqq{ Q= {2\k^2_{D-1} \ov (D-4) \om_{D-3} } \cdot {2\pi N\ov L}=
     {\k^2_{4} N\ov L} =   {\k^2 N\ov L^8 }\ , }
where $\k_4^2/8 \pi  $ and $\k^2/8\pi $ 
are the Newton's constants in 4 and 11 dimensions.
All toroidal directions are assumed to have length $L$.

%%%%%%%%%%%%%%%%%%%%%%%%%%%%%%%%%%%%%%%%%%%%%%%%%%
The three magnetic charges are proportional to the
numbers  $n_1,n_2,n_3$  of 5-branes in the $(14567)$,   $(12367)$, and the
$(12345)$  planes, respectively (see \fivv,\fivve). 
The complete symmetry between
$n_1$, $n_2$ and $n_3$ is thus 
automatic in the 11-dimensional approach. 
The precise relation between $P_i$ and $n_i$ is found as follows.
The charge $q_5$ of a $D=11$ 5-brane  which is spherically symmetric 
in transverse $d+2 \leq 5$ dimensions is proportional to  the coefficient $P$ 
in the corresponding harmonic function. 
For $d+2=3$ appropriate to the present case (two of five  transverse 
directions are isotropic, or, equivalently,  there is 
 a periodic array 
of 5-branes in these compact directions)  we get  
\eqn\qdq{  q_5= {\om_{d+1} d \ov \sqrt 2 \k } P  \ \rightarrow \   
{\om_{2} L^2 \ov \sqrt 2 \k } P = {4\pi  L^2 \ov \sqrt 2 \k } P  \ . } 

At this point we need to know precisely how the 5-brane charge is
quantized. This was discussed in \KT, but we repeat the argument here
for completeness. 
A different argument leading to equivalent results was presented earlier
in \jons. Upon compactification on a circle of length $L$,
the M-theory reduces to type IIA string theory where all charge quantization
rules are known. We use the fact that double dimensional reduction
turns a 2-brane into a fundamental string, and a 5-brane into a 
Dirichlet 4-brane. Hence, we have
\eqn\doubled{
T_2 \k^2 = T_1 \k^2_{10} \ ,\qquad\qquad T_5 \k^2 = T_4 \k^2_{10} \ ,}
where
the 10-dimensional gravitational constant is expressed in terms of
the 11-dimensional one by $\k^2_{10}= \k^2/L$.
The charge densities are related to the tensions by
\eqn\yry{
q_2  = \sqrt 2 \k T_2 \ ,\qquad q_5  = \sqrt 2 \k T_5\ ,}
and we assume that the minimal Dirac condition is satisfied,
$q_2 q_5= 2\pi$. These relations, together with the 10-dimensional
expressions \jch
\eqn\tendim{
\k_{10}=g (\alpha')^2\ ,\qquad T_1={1\over 2\pi\alpha'}\ ,
\qquad \k_{10} T_4= {1\over 2\sqrt{\pi\alpha'} } \ ,}
fix all the M-theory quantities in terms of $\alpha'$
and the string coupling constant, $g$.
In particular, we find
\eqn\plan{\k^2= {g^3 (\alpha')^{9/2}\over 4 \pi^{5/2} }\ ,
\qquad L= {g \sqrt {\alpha'} \over 4 \pi^{5/2}} \ .}
The tensions turn out to be
\eqn\tensions{ T_2={2\pi^{3/2}\over g (\alpha')^{3/2}}\ ,\qquad
T_5= {2\pi^2\over g^2 (\alpha')^3} \ .}
Note that $T_2$ is identical to the tension of the Dirichlet 2-brane
of type IIA theory, while $T_5$ -- to the tension of the solitonic
5-brane. 
This provides a nice check on our results, since single dimensional
reduction indeed turns the M-theory 2-brane into the Dirichlet 2-brane,
and the M-theory 5-brane into the solitonic 5-brane.
Note that the M-brane tensions satisfy the relation
$2\pi T_5 =  T_2^2$, which was first derived in \jons\ using
toroidal compactification to type IIB theory in 9 dimensions.
This serves as yet another consistency check.

It is convenient to express our results in pure M-theory terms.
The charges are quantized according to\foot{In
\mina\ it was argued that
the 2-brane tension, $T_2$, satisfies  $\k^2 T^3_2 = \pi^2/m_0$,
where $m_0$ is a rational number that was left undetermined.
The argument of \jons, as well as our procedure \KT, 
unambiguously fix $m_0=1/2$.}
\eqn\quaa{
q_2 =\sqrt 2 \k T_2 =   n \sqrt 2 (2\k \pi^2)^{1/3} \ , }
 \eqn\five{  q_5 = \sqrt 2 \k T_5=  n \sqrt 2 ({\pi\ov 2\k})^{1/3} \ ,
}
i.e.
\eqn\fivee{
P_i= { n_i \ov 2 \pi L^2} ({\pi \k^2 \ov 2})^{1/3} \ . }

The resulting expression for the 
Bekenstein-Hawking entropy of the extremal Reissner-Nordstr\"om type 
black hole, \ddd, \luo, which is proportional to the 
area \are,  is 
\eqn\exen{ S_{BH} = {2\pi A_9 \ov \k^2} 
={8\pi^2 L^7  \ov \k^2}\sqrt {P_1P_2P_3 Q}
 = 2\pi \sqrt { n_1n_2n_3 N }  \ .  }
This agrees with the expression found directly in $D=4$ 
\refs{\LW,\CTT,\mast,\myers}. 

In the case of the $2\bot2\bot5\bot 5$
configuration  we find (for each pair of 2-brane and 5-brane charges)
$q_2 = {4\pi  L^5 \ov \sqrt 2 \k } Q, \ 
q_5 = {4\pi  L^2 \ov \sqrt 2 \k } P.$ 
The Dirac condition on unit charges
translates into  
 $ q_2q_5 = 2\pi n_1 n_2 $, where $n_1$ and $n_2$ are the numbers
of 2- and 5-branes. We conclude that 
$Q_1 P_1 = {\k^2 \ov 4\pi L^7} n_1n_2$. Then from 
\area\  we learn that 
\eqn\arre{  S_{BH} = {2\pi A_9 \ov \k^2} 
={8\pi^2 L^7  \ov \k^2}\sqrt {Q_1 P_1 Q_2 P_2 }
 = 2\pi \sqrt { n_1n_2n_3 n_4 }  \ .}
Remarkably, this result does not depend on the particular 
choice of M-brane quantization condition 
(choice of $m_0=\pi^2 \k^{-2} T_2^{-3}$) or  use of D-brane tension expression
since the $2\bot2\bot5\bot 5$
 configuration contains equal number of 2-branes and 5-branes.
This provides a consistency check. Note also that the $D=4$ black holes
obtained from the $2\bot2\bot5\bot 5$ and from the $5\bot5\bot 5$
M-theory configurations are not identical, but are related by U-duality.
The equality of their entropies provides a check of the U-duality.

The same  expression is obtained  for the entropy 
of the   $D=10$ configuration $3\bot3\bot3\bot 3$  \fqw\  
(or related $D=4$ black hole).
Each 3-brane charge  $q_{3}$ is proportional   to the corresponding
 coefficient $Q$ 
in the harmonic function (cf. \qdq)
\eqn\ouou{
q_3= {1 \ov \sqrt 2} ({\om_{d+1} d \ov
 \sqrt 2 \k_{10} })  Q  \  \ \rightarrow  \   \
{\om_{2} L^3 \ov  2 \k_{10} } Q = {2\pi  L^3 \ov   \k_{10} } Q\ ,  }
where $\k_{10}^2/8\pi$ is the 10-dimensional Newton's constant and 
the overall factor $ {1 \ov \sqrt 2}$ is due  to the dyonic
nature of the 3-brane. The charge quantization in the self-dual case
implies (see \KT)  $ q_3 =  n \sqrt \pi $ (the absence of standard 
$\sqrt 2$ factor here effectively compensates for the `dyonic'
${1 \ov \sqrt 2}$ factor in the expression for the charge).\foot{This agrees
with the D3-brane tension,   $\k_{10} T_3=\sqrt \pi$,
since in the self-dual case   $q_p = \k_{10}  T_p$.} 
Thus, $Q_i= {\k_{10}  \ov 2\sqrt \pi} n_i$, and 
 the area \arr\  leads to the following entropy,
\eqn\ioio{ S_{BH} = {2\pi A_8 \ov \k_{10}^2} 
={8\pi^2 L^6  \ov \k^2}\sqrt {Q_1Q_2Q_3 Q_4}
 = 2\pi \sqrt { n_1n_2n_3 n_4 }\ .}
%%%%%%%%%%%%%%%%%%%%%%%%%%%%%%%%%%%%%%%%%%%%%%%

\subsec{Counting of the microscopic states}
%%%%%%%%%%%%%%%%%%%%%%%%%%%%%%%%%%%%%%%%%%%%%
The presence of the factor $\sqrt N$ in $S_{BH}$ \exen\ 
immediately suggests an
interpretation in terms of the massless states on the string common to
all three  5-branes.
Indeed, it is well-known that, for a $1+1$ dimensional field theory
with a central charge $c$, the entropy of left-moving states
with momentum $2\pi N/L$ is, for sufficiently large $N$, given by\foot{
As pointed out in \MalLen, this expression is reliable only
if $N\gg c$. Requiring $N$ to be much greater than $n_1 n_2 n_3$
is a highly asymmetric choice of charges. If, however, all charges
are comparable and large, the entropy is dominated by the multiply
wound 5-branes, which we discuss at the end of this section.} 
\eqn\uuu{ S_{stat} = 2\pi \sqrt{{\textstyle {1 \over 6}} cN} \ . }
We should find, therefore, that the central charge on the
intersection string is, in the limit of large charges, equal to
\eqn\iii{  c = 6 n_1 n_2 n_3 \ . } 
The fact that the central charge grows as 
$n_1 n_2 n_3$ suggests the following picture. 
 2-branes can end on 5-branes, so that the boundary looks like a closed 
string \refs{\ast,\ttt,\beck}. 
It is tempting to associate the massless states with those
of  2-branes attached to 5-branes near the intersection point.
Geometrically, we may have a two-brane with three holes, each of the
holes attached to different 5-dimensional hyperplanes in which the
5-branes lie.
Thus, for any three 5-branes that intersect along a line, we have
a collapsed 2-brane that gives massless states
in the $1+1$ dimensional theory describing the intersection.
What is the central charge
 of these massless states? From the point of view of one of the 5-branes,
the intersection is a long string in $5+1$ dimensions.
Such a string has 4 bosonic massless modes corresponding to
the transverse oscillations, and 4 fermionic superpartners.
Thus, we believe that the central charge arising from the
collapsed 2-brane with three boundaries is
$4(1 + \ha)= 6$.\foot{Upon compactification on $T^7$, these massless
modes are simply the small fluctuations of the long string in
$4+1$ dimensions which is described by the classical solution
\soo. One should be able to confirm that the central charge on this
string is equal to 6 by studying its low-energy modes.}

 The upshot of this
argument is that each triple intersection contributes $6$ to the
central charge. Since there are $n_1 n_2 n_3$ triple intersections,
we find the total central charge $6 n_1 n_2 n_3$.
One may ask why there are no terms of order $n_1^3$, etc. This can be
explained by the fact that all parallel 5-branes are displaced relative
to each other, so that the 2-branes produce massless states only near
the intersection points.

One notable feature of our argument is that the central charge grows
as a product of three charges, while in all D-brane examples one
found only a product of two charges. We believe that this is related
to the peculiar $n^3$ growth of the near-extremal entropy
of $n$ coincident 5-branes
found in \KT\  (for coincident D-branes the near-extremal entropy grows
only as $n^2$). This is because the intersecting
D-brane entropy comes from strings which can only connect objects
pairwise. The 2-branes, however, can connect three different 5-branes.
Based on our observations about entropy, we conjecture that the
geometries where a 2-brane connects four or more 5-branes are 
forbidden (otherwise, for instance, the near-extremal entropy 
of $n$ parallel 5-branes would grow faster than $n^3$).
Perhaps such configurations are not supersymmetric and do not give
rise to massless states.

The counting argument presented above applies to the configuration where
there are $n_1$ parallel 5-branes in the $(14567)$ hyperplane,
$n_2$ parallel 5-branes in the $(12367)$ hyperplane, and
$n_3$ parallel 5-branes in the $(12345)$ hyperplane.
As explained in \MalLen, if $n_1\sim n_2\sim n_3 \sim N$
we need to examine a different
configuration where one replaces a number of disconnected branes by
a single multiply wound brane. Let us consider, therefore,
a single 5-brane in the $(14567)$ hyperplane wound $n_1$ times
around the $y_1$-circle, 
a single 5-brane in the $(12367)$ hyperplane wound $n_2$ times
around the $y_1$-circle, and
a single 5-brane in the $(12345)$ hyperplane wound $n_3$ times
around the $y_1$-circle. Following the logic of \MalLen, one can show
that the intersection string effectively
has winding number $n_1 n_2 n_3$: this is because the 2-brane which
connects the three 5-branes needs to be transported
$n_1 n_2 n_3$ times
around the $y_1$-circle to come back to its original state.\foot{The 
 role of  $n_1 n_2 n_3$ as the effective winding number 
is  suggested also by comparison of 
the $D=5$ solitonic string metric, \soo,   with
the fundamental string metric, 
$ds^2= V\inv  (dudv  + Kdu^2)+ dx_sdx_s$ ,
where the coefficient in the 
harmonic function $V$ is proportional to the tension times the winding number
of the source string  (see e.g. \dabb).
After a conformal rescaling, \soo\ takes the fundamental string form with 
$V=H^3 = (F_1F_2F_3)\inv$ so that 
 near $r=0$ the $dudv$ part of it 
is multiplied  by $P_1P_2P_3 \sim n_1n_2n_3$.
Thus, the source string may
be thought of as wound $n_1n_2n_3$ times around the
circle.}
Therefore, the massless fields produced by the 2-brane effectively live
on a circle of length $n_1 n_2 n_3 L$.
This implies \DasMat\ that the energy levels
of the $1+1$ dimensional field theory are quantized in units of
$2\pi/(n_1 n_2 n_3 L)$. In this theory there is only one species
of the 2-brane connecting the three 5-branes; therefore, the central
charge on the string is $c=6$. The calculation of BPS
entropy for a state with momentum $2\pi N/L$, as in
\refs{\DasMat,\MalLen}, once again reproduces \exen.
While the end result has the form
identical to that found for the disconnected
5-branes, the connected configuration is dominant when all four charges
are of comparable magnitude \MalLen. Now
the central charge is fixed, and
the large entropy is due to the growing density of energy levels.

%%%%%%%%%%%%%%%%%%%%%%%%%%%%%%%%%%%%%%%%
\newsec{Black Hole Entropy in $D=5$ and Discussion.}
%%%%%%%%%%%%%%%%%%%%%%%%%%%%%%%%%%%%%%%%%
The counting arguments
presented here are plausible, but clearly need to be put
on a more solid footing. 
Indeed, it is not yet completely clear what rules apply to
the 11-dimensional M-theory (although 
 progress has been
made in \dvv). The rule associating massless states to collapsed
2-branes with three boundaries  looks  natural, and seems to reproduce
the Bekenstein-Hawking entropy of extremal black holes in $D=4$.
Note also that a similar rule 
can be successfully applied  to 
 the case of the  finite entropy $D=5$ extremal dyonic black holes 
described in 11 dimensions by the 
`boosted' $2\perp 5$ configuration \AT. Another possible $D=11$ embedding 
of the $D=5$ black hole is 
provided by $2\bot2\bot2$ configuration \AT. The relevant $D=10$ type IIB 
configuration is $3\bot 3$ (cf. \fqw)
with momentum flow along common string.
In the case of $2\perp 5$ configuration 
the massless degrees of freedom on the intersection string may
be attributed to a collapsed 2-brane with a hole attached to the 5-brane
and one point attached to the 2-brane. If the 5-brane is wound $n_1$
times and the 2-brane -- $n_2$ times, the intersection is described
by a $c=6$ theory on a
circle of length $n_1 n_2 L$. Following
the arguments of \MalLen, we find that the entropy of a state with
momentum $2\pi N/L$ along the intersection string is
\eqn\fiven{S_{stat} =2\pi\sqrt{n_1 n_2 N}\ . }
This seems to supply a microscopic M-theory basis, somewhat different
from that in \dvv, for the
Bekenstein-Hawking entropy of $D=5$ extremal dyonic
 black holes.

We would now like to show that \fiven\ is  indeed 
equal to the expression 
for the Bekenstein-Hawking 
entropy for  the `boosted' $2\perp 5$ configuration  \AT\  (cf. \exen)
\eqn\boosted{S_{BH} = {2\pi A_9 \ov \k^2} 
     ={4\pi^3  L^6  \ov \k^2}\sqrt {Q P Q'} \ .}
$Q$ and $P$ are the parameters in the harmonic
functions corresponding to the 2-brane and the 5-brane,
and $Q'$ is the  parameter in the `boost' function, 
i.e. $T\inv = 1 + Q/r^2$, $F\inv = 1 + P/r^2$,  $K=1 + Q'/r^2$. 
Note that  here (cf. \qqq)
\eqn\opi{Q' = {\k^2 N\ov \pi L^7 }\ ,\qquad
q_2 = {4\pi^2  L^4 \ov \sqrt 2 \k } Q \ ,\qquad 
q_5 = {4\pi^2  L \ov \sqrt 2 \k } P \ .}
As in the case
of the $2\bot 2\bot 5\bot 5$ configuration, 
we can use the Dirac  quantization condition, 
$ q_2q_5 = 2\pi n_1 n_2 $, to conclude
 that $QP = {\k^2 \ov  4\pi^3 L^5} n_1n_2$. This yields \fiven\ when
substituted into \boosted.
A similar expression for the BPS entropy
is found in the case of the completely symmetric
$2\bot2\bot2$
configuration, 
\eqn\entrq{S_{BH} 
={4\pi^3  L^6  \ov \k^2}\sqrt {Q_1Q_2 Q_3}=2\pi\sqrt{n_1 n_2 n_3}\ , 
}
where we have used the 2-brane charge quantization condition
\quaa, which implies that $Q_i = n_i L^{-4} ({\k\ov \sqrt 2  \pi})^{4/3}$.
Agreement of different expressions for the $D=5$ black hole 
entropy provides another check on the consistency of \quaa,\five.

Our arguments for   counting the microscopic states applies only to
the configurations where M-branes intersect over a string.
It would be very interesting to see how approach 
analogous to the above might work when this is not the case. 
Indeed, black holes with finite horizon area in $D=4$
may also be obtained from the 
$2\bot 2\bot 5\bot 5$  configuration  in M-theory,
and the $3\bot 3\bot 3\bot 3$ one in type IIB, while in
$D=5$ -- from the $2\bot2\bot2$ configuration.
Although from the $D=4,5$ dimensional point of view
these cases are related by U-duality to the ones we considered,
the counting of their states 
seems to be harder at the present level of understanding.
We hope that a more general approach to the entropy problem, which covers
all the solutions we discussed, can be found.

%%%%%%%%%%%%%%%%%%%%%%%%

%%%%%%%%%%%%%%%%%%%%%%%%%%%%%%%%%%%%%%%%
\newsec{ Acknowledgements}
%%%%%%%%%%%%%%%%%%%%
We are grateful to  V. Balasubramanian, C. Callan and 
M. Cveti\v c for useful discussions.
I.R.K. was supported in part by DOE grant DE-FG02-91ER40671, the NSF
Presidential Young Investigator Award PHY-9157482, and the James S.{}
McDonnell Foundation grant No.{} 91-48.
A.A.T. would like to 
acknowledge  the support of PPARC,
ECC grant SC1$^*$-CT92-0789 and NATO grant CRG 940870.

\vfill\eject
\listrefs
\end